\documentclass[floatfix,prl,twocolumn,showpacs,preprintnumbers,amsmath,amssymb]{revtex4}
\usepackage{graphicx}

\begin{document}
%\draft
%\twocolumn[\hsize\textwidth\columnwidth\hsize
%           \csname @twocolumnfalse\endcsname

%\title{Quantum molecular dynamics simulations for dense helium}
%\title{Nonmetal-to-metal transition in fluid helium: a QMD study}
\title{Quantum molecular dynamics simulations for the nonmetal-to-metal transition in fluid helium}

\author{Andr\'e Kietzmann, Bastian Holst, and Ronald Redmer}
\affiliation{Universit\"at Rostock, Institut f\"ur Physik, 
 D-18051 Rostock, Germany}

\author{Michael P. Desjarlais and Thomas R. Mattsson}
\affiliation{Pulsed Power Science Center, Sandia National Laboratories,
  Albuquerque, New Mexico 87185-1186, USA}

%\author{Vladimir Ya. Ternovoi}
%\affiliation{Institute for Chemical Physics Research, Russian Academy of Sciences, 
%  Chernogolovka, 142432, Russia} 

\date{\today}

\begin{abstract}
  We have performed quantum molecular dynamics simulations for dense helium
  to study the nonmetal-to-metal transition at high pressures. We present 
  new results for the equation of state and the Hugoniot curve in the warm 
  dense matter region. The optical conductivity is calculated via the
  Kubo-Greenwood formula from which the dc conductivity 
  is derived. The nonmetal-to-metal transition is identified at about
  1~g/cm$^3$. We compare with experimental results as well as with other
  theoretical approaches, especially with predictions of chemical models.
\end{abstract}
\vspace*{5mm}
%]

\maketitle

%%%%%%%%%%%%%%%%%%%%%%%%%%%%%%%% Introduction 
%\section{Introduction}

Hydrogen and helium are by far the most abundant elements in nature. The
investigation of their phase diagram, especially at extreme conditions of
pressure and temperature, is not only of fundamental interest but also an
indispensable prerequisite for astrophysics. For instance, thermophysical
properties of hydrogen-helium mixtures determine the structure and evolution
of stars, White Dwarfs, and Giant Planets~\cite{SCVH,TernoH-He,Vor+06}. The
detection of Jupiter-like planets orbiting neighboring stars~\cite{planets}
has initiated a renewed interest in planetary physics. All planetary models
require an input of accurate equation of state data in order to solve the
hydrostatic equations for a rotating compact object. While the limiting cases
of low and high densities are well understood within chemical and plasma
models, the intermediate region is much more complex. There, a
nonmetal-to-metal transition occurs in both hydrogen and helium at pressures
of several megabar and temperatures of few eV which implies a strong
state-dependence of the interparticle interactions and, thus, also of the
thermodynamic variables. In this paper, we study this
{\it warm dense matter} region where the uncertainties in the equation of
state data, both experimentally and theoretically, are greatest.

A lot of effort has been done to understand the behavior of warm dense
hydrogen~\cite{HPR2000,Nellis06}.  Although the simplest
element in the periodic table, the transition of a nonconducting molecular
liquid to an atomic or plasma-like conducting fluid at high pressure is still
not fully understood~\cite{Ashcroft,LAC,Hopping}. Helium seems to be a much
simpler system for the study of the high-pressure behavior since no
dissociation equilibrium between molecules and atoms interferes with the
ionization equilibrium, and the first (24.6~eV) and second ionization energy
(54.4~eV) are well separated.  Surprisingly, only few experimental and
theoretical studies exist for warm dense helium~\cite{EbelHe,Schlanges,Winis}.

In this paper, we perform the first comprehensive {\it ab initio} study of the
high-pressure behavior of helium. We
determine the equation of state (EOS) in the warm dense matter region by means
of quantum molecular dynamics (QMD) simulations. The Hugoniot curve is derived
and compared with experimental points~\cite{Nellis84}, other {\it ab initio} 
calculations~\cite{Mil06}, as well as with results of efficient chemical
models~\cite{EbelHe,Schlanges,Winis}. Finally, we calculate the dynamic
conductivity via the Kubo-Greenwood formula and derive the static conductivity
and compare with shock wave experimental data~\cite{TernoHeSig}. We locate the 
nonmetal-to-metal transition in the high-pressure phase diagram and discuss 
the related plasma phase transition (PPT).

%%%%%%%%%%%%%%%%%%%%%%%%%%%%%%%% QMD
%\section{Quantum molecular dynamics simulations}

QMD simulations are a powerful tool to calculate the structural,
thermodynamic, and optical properties of warm dense matter in a combined
first-principles approach. Details of this method have been described
elsewhere~\cite{MPD+02,MPD03,LAU+04,CLE+05}.

We have performed {\it ab initio} molecular dynamics simulations 
employing a finite temperature Fermi occupation of the electronic 
states using Mermins finite temperature density functional
theory (FT-DFT)~\cite{Mermin65}. The implementation of the QMD method
comes from VASP (Vienna Ab Initio Simulation Package), a plane wave
density functional code developed at the University of
Vienna~\cite{VASP1}. All electrons are treated 
quantum mechanically on the level of DFT. The
electronic wave function is relaxed at each QMD step, which assumes
decoupled electronic and ionic time scales. We have chosen a
simulation box with 32 to 64~atoms and periodic boundary conditions. 
The electron wave functions are modeled using the projector augmented 
wave (PAW) potentials~\cite{PAW} supplied with VASP~\cite{VASP1}. 
These PAW potentials enable more accurate results for conductivity 
calculations compared with other pseudopotentials. The exchange 
correlation functionals are
calculated within generalized gradient approximation (GGA). Our most 
accurate calculations were done using the GGA parameterization of 
PBE~\cite{PBE}. The convergence of the thermodynamic quantities in 
QMD simulations is of significant importance~\cite{MATS05}. We have 
chosen a plane wave cutoff $E_{\rm cut}$ at 700~eV where the pressure 
is converged to within 2\%.

Performing QMD simulations to calculate the EOS of He only the $\Gamma$ 
point was used for the representation of the Brillouin zone. 
Calculations for Al have shown that it is not recommended to
calculate higher-order $\textbf{k}$-point sets~\cite{MPD+02}. Furthermore, 
the mean value point (1/4, 1/4, 1/4) was used for conductivity 
calculations.

Our simulations were performed for a canonical ensemble where the 
temperature, the volume of the simulation box, and the particle number 
in the box are conserved quantities during a QMD run.
To keep the temperature on a predefined level, the ion temperature is 
regulated by a Nos\'{e}-Hoover thermostat and the electronic temperature 
is fixed by Fermi weighting the occupation of bands~\cite{VASP1}. 
After about hundred time steps the system is equilibrated and the 
subsequent 400 to 1000~time steps are taken to calculate pressures, 
energies and other quantities as averages over this simulation period.

%%%%%%%%%%%%%%%%%%%%%%%%%%%%%%%% EOS
%\section{Equation of state}

\begin{figure}[ht]
\center{\includegraphics*[angle=0,width=7cm]{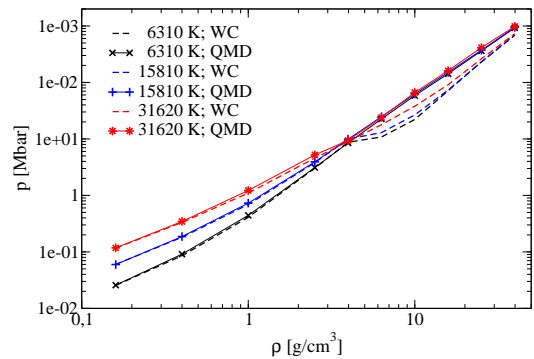}}
\caption{Pressure isotherms in comparison with the WC free energy 
 model~\protect\cite{Winis}. 
\label{P}} 
\end{figure}

\begin{figure}[ht]
\center{\includegraphics*[angle=0,width=7cm]{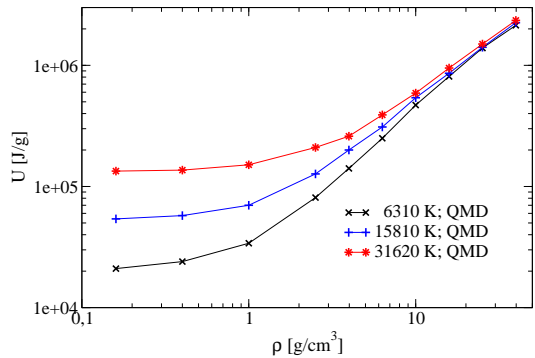}}
\caption{Isotherms of the internal energy. 
\label{U}}
\end{figure}

First we present results for the thermal and caloric EOS of warm dense helium 
in Figs.~\ref{P} and \ref{U}. The isotherms of the pressure and the internal
energy behave very systematically with temperature and density and show no 
indications of an instability such as the PPT at lower temperatures, contrary 
to results derived within the chemical picture~\cite{EbelHe,Schlanges,Winis}. 
For instance, the EOS of Winisdoerffer and Chabrier (WC)~\cite{Winis} is 
based on a free energy minimization schema for a mixture of helium atoms, 
single and double charged ions, and free electrons. Correlations are taken 
into account based on effective two-particle potentials. It agrees well with 
our QMD results for the pressure up to about 1~g/cm$^3$ and for ultra-high 
densities above about 50~g/cm$^3$. However, this chemical model shows a 
systematic trend to lower pressures in the intermediate, strongly coupled 
region where the QMD results already approach an almost temperature-independent 
behavior as characteristic of a degenerate electron gas. These results underline 
the significance of {\it ab initio} calculations for warm dense matter states 
and will have a strong impact on calculations of planetary 
interiors~\cite{ApJ2004}. Furthermore, we can identify the region where 
efficient chemical models are applicable in favor of time-consuming 
{\it ab initio} calculations. 

% Fig 1 Isotherms P(rho)

% Fig 2 Isotherms U(rho)

\begin{figure}[ht]
\center{\includegraphics*[angle=0,width=7cm]{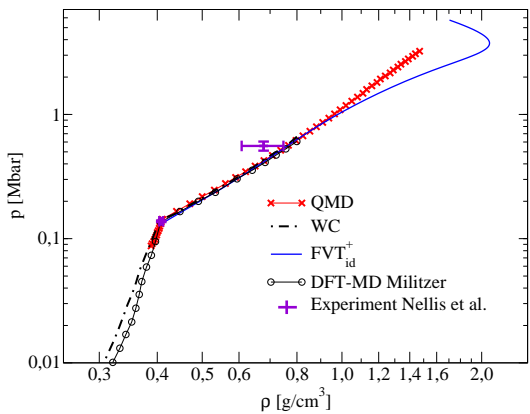}}
\caption{Hugoniot curve for helium: QMD results are compared with double-shock
  experiments of Nellis {\it et al.}~\protect\cite{Nellis84}, zero-Kelvin 
  DFT-MD results of Militzer~\protect\cite{Mil06} accounting for excited states, 
  and two chemical models WC~\protect\cite{Winis} and FVT$^+_{\rm id}$~\cite{PCCP05}.
\label{Hugo}} 
\end{figure}

Based on this EOS data, we have determined the Hugoniot curve via the condition
\begin{align}
(E-E_0)=\frac{1}{2}(P+P_0)(V_0-V),
\end{align}
which relates all final states of a shock wave experiment $(E,P,V)$ with the 
initial state $(E_0,P_0,V_0)$. In our calculations we have used the values 
$E_0=20$~J/g, $T_0=4$~K, $P_0=1$~bar$\ll P$, and $V_0=32.4$~cm$^3$/mol 
($\varrho_0=0.1235$~g/cm$^3$) for the first shock and $E_1=57$~kJ/g, 
$T_1=15.000$~K, $P_1=190$~kbar, and $V_1=9.76$~cm$^3$/mol 
($\varrho_1=0.41$~g/cm$^3$) for the secound shock. 
We compare our results with double-shock 
experiments of Nellis {\it et al.}~\cite{Nellis84}, 
with recent DFT-MD calculations of Militzer~\cite{Mil06}, 
and with two chemical models~\cite{Winis,PCCP05} in Fig.~\ref{Hugo}. 

A very good agreement is found with the double-shock experiments as well as with 
the other theoretical Hugoniot curves up to about 1~g/cm$^3$ which results from 
the accordance of the EOS data up to this density as mentioned above. A central  
problem in this context is the value and location of the maximum compression 
ratio $\eta_{\rm max}$. The chemical model FVT$^+_{\rm id}$ 
%of Schwarz {\it et al.}~
\cite{PCCP05} is based on fluid variational theory and considers an 
ideal ionization equilibrium in addition. This model predicts a value of 5.5 at 
375~GPa and 50.000~K if the second shock of the experiment is taken as initial 
state. Militzer~\cite{Mil06} found a maximum compression of 5.24 at 360~GPa and 
150.000~K for the principal Hugoniot starting at $(E_0,P_0,V_0)$ by using an 
EOS composed of zero-Kelvin DFT-MD results accounting for excited states for 
lower temperatures as shown in Fig.~\ref{Hugo} and Path Integral Monte Carlo 
(PIMC) data in the high-temperature limit. 
Note that only a finite-temperature DFT calculation
yields the self-consistent thermal ground state of the system, which is not
equivalent to applying a thermal occupation of the empty electronic states
(``excited states'') obtained in a zero-Kelvin electronic structure
calculation as performed in~\cite{Mil06}; for details, see~\cite{MattDesjH2O}.

The QMD simulations were performed up to 1.5~g/cm$^3$, 350~GPa, and 60.000~K 
where the maximum compression ratio has not been reached yet. For still higher 
temperatures the number of bands increases drastically beyond the scope of our 
computer capacity. Besides PIMC simulations~\cite{Mil06}, orbital-free DFT 
methods may be applicable in that high-temperature region~\cite{OF-DFT}.
Interestingly, the maximum compression ratio for helium ($\eta_{\rm max}\ge5$) 
is greater than that for hydrogen ($\eta_{\rm max}=4.25$); see 
e.g.~\cite{Nellis06} for a more detailed discussion.

% Fig 3 Hugoniot curve

%%%%%%%%%%%%%%%%%%%%%%%%%%%%%% Conductivity
%\section{Dynamic conductivity}

The dynamic conductivity $\sigma_{\mathbf k}(\omega)$ for one ${\mathbf k}$ 
point is derived from the Kubo-Greenwood formula~\cite{Kubo,Greenwood}
\begin{align}\label{KG}
\sigma_{\bf k}(\omega) = \frac{2\pi e^2\hbar^2}{3 m^2 \omega \Omega}
\sum_{j=1}^N\sum_{i=1}^N\sum_{\alpha=1}^3 
\left[ F(\epsilon_{i,{\mathbf k}})-F(\epsilon_{j,{\mathbf k}})\right] \notag\\
\times |\langle\Psi_{j,{\mathbf k}}|\nabla_\alpha|\Psi_{i,{\mathbf k}}\rangle|^2
\delta(\epsilon_{j,{\mathbf k}}-\epsilon_{i,{\mathbf k}}-\hbar \omega),
\end{align}
where $e$ is the electron charge and $m$ its mass. The summations over $i$ 
and $j$ run over $N$ descrete bands considered in the electronic structure 
calculation for the cubic supercell volume $\Omega$. The three spatial 
directions are averaged by the $\alpha$ sum.
$F(\epsilon_{i,{\mathbf k}})$ describes the occupation of the $i$th band
corresponding to to the energy $\epsilon_{i,{\mathbf k}}$ and the wavefunction
$\Psi_{i,{\mathbf k}}$ at ${\bf k}$. $\delta$ must be broadened because of the
discrete eigenvalues resulting from the finite simulation volume~\cite{MPD+02}. 
Integration over the Brillouin zone is done by sampling special ${\mathbf k}$ 
points~\cite{Monkhorst},
\begin{equation}
\sigma(\omega)=\sum_{\mathbf k} \sigma_{\mathbf k} (\omega)W({\mathbf k}),
\end{equation}
where $W({\mathbf k})$ is the weighting factor for the respective ${\mathbf k}$
point. Calculations are usually done at the mean value point~\cite{Baldereschi}.

% Fig 4 DC conductivity
\begin{figure}[ht]
\center{\includegraphics*[angle=0,width=7cm]{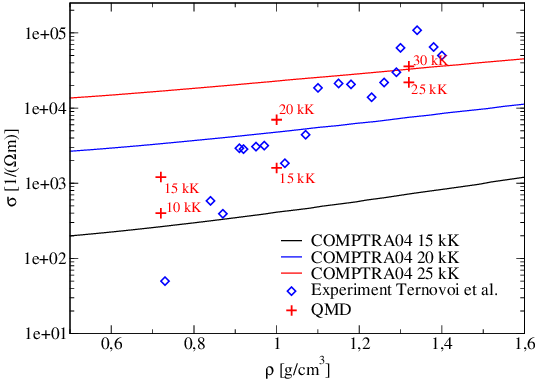}}
\caption{QMD results for the static conductivity are compared with shock-wave 
 experiments of Ternovoi {\it et al.}~\protect\cite{TernoHeSig} between 
 $(15-25)\times10^3$~K and isotherms of the COMPTRA04 model~\cite{KUH+05}; 
 temperatures are indicated. \label{sigma-dc}}
\end{figure}

The dc conductivity follows from Eq.~(\ref{KG}) in the limit $\omega\to 0$.
We compare with isentropic compression experiments of 
Ternovoi~{\it et al.}~\cite{TernoHeSig} performed in the range 
$(15-25)\times10^3$~K and predictions of the partially ionized plasma model COMPTRA04~\cite{Comptra04,KUH+05} in Fig.~\ref{sigma-dc}. 
The experimental points show a very strong 
increase between 0.7 and 1.4~g/cm$^3$ indicating that a nonmetal-to-metal 
transition occurs. Using the Mott criterion for the minimum metallic 
conductivity of $20000/\Omega$m also for finite temperatures, this transition 
can be located at about 1.3~g/cm$^3$. The QMD results reproduce the strong 
increase found experimentally very well except for the lowest density of 
0.72~g/cm$^3$ where the experimental value is substantially lower than the 
QMD result. 

The COMPTRA04 model~\cite{Comptra04,KUH+05} calculates the ionization degree and, 
simultaneously, the electrical conductivity accounting for all scattering processes 
of free electrons in a partially ionized plasma. This approach is able to describe
the general trends of the electrical conductivity with the density and temperature 
as found experimentally, see~\cite{Fortov03}. The isotherms displayed in 
Fig.~\ref{sigma-dc} cover almost the range of the experimental points and agree 
also with the QMD data so that the nonmetal-to-metal transiton is described 
qualitatively very well.

The strong influence of the temperature on the dc conductivity in this 
transition region can be seen by comparing the QMD results for two 
temperatures at the same density point; see Fig.~\ref{sigma-dc}. In order to 
exclude systematic errors from the experimental temperature determination, 
we have performed EOS calculations for the experimental points and compare 
typical values in Table~I. A very good agreement is obtained so 
that the discrepancy in the conductivity data for the lowest density stems 
probably from the band gap problem of DFT. In order to solve this problem, DFT 
calculations beyond the GGA have to be performed by using, e.g., exact exchange 
formalsims~\cite{EXX} or quasi-particle calculations~\cite{GW}. This is an 
important issue of future work devoted to the nonmetal-to-metal transition.

\begin{table}[ht]
\caption{Comparison of EOS data inferred from the experiment~\protect\cite{TernoHeSig} 
 and QMD data.} 
\begin{tabular}{cccc}\hline\hline
$\rho$ [g/cm$^{3}]$ & $T$ [K] & $p_{\rm Exp}$ [kbar] & $p_{\rm QMD}$ [kbar]\\ \hline 
0.73 & 16700 & 500  & 478 \\ 
1.02 & 19400 & 900  & 892 \\ 
1.38 & 23500 & 1685 & 1639 \\ \hline\hline
%1.15 & 15800 & 940  & x  \\ \hline
%1.29 & 14400 & 1100  & x  \\ \hline
\end{tabular}
\end{table}

\begin{figure}[ht]
\center{\includegraphics*[angle=0,width=7cm]{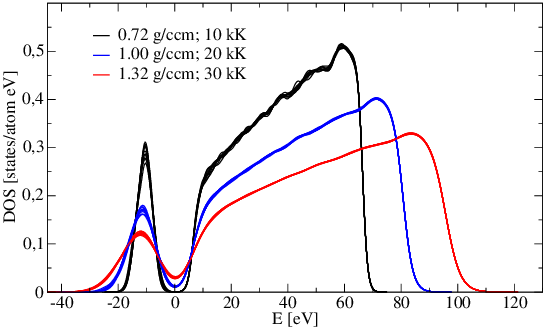}}
\caption{DOS as function of the energy for three typical situations. 
 The Fermi energy is located at zero.
\label{fig-DOS}}
\end{figure}

The origin of this nonmetal-to-metal transition can be elucidated by inspecting 
the variation of the density of states (DOS) along the path of the shock-wave 
experiments, see Fig.~\ref{fig-DOS}. 
For the lowest density of 0.72~g/cm$^3$, a gap still exists in the DOS so that 
thermally activated electron transport occurs as in semiconductors. With 
increasing density, the gap region is slightly reduced. The main effect is, 
however, that electronic states fill up the region of the Fermi energy with 
increasing temperature so that a higher, metal-like conductivity follows, 
see also~\cite{LAC}.

% Fig 5 DOS

%%%%%%%%%%%%%%%%%%%%%%%%%%%%%%% Conclusions
%\section{Conclusion}

In summary, we have determined the thermophysical properties of dense helium 
within an {\it ab initio} approach. The results % for EOS and the Hugoniot curve 
show clearly the strong influence of quantum effects and correlations in the 
warm dense matter region. The nonmetal-to-metal transition occurs at about 
1~g/cm$^3$, in good agreement with shock wave experimental data. 
These new results will have a strong influence on models for planetary interiors.

%%%%%%%%%%%%%%%%%%%%%%%%%%%%%%% Acknowledgements.
%\begin{acknowledgements}
  We thank P.M.~Celliers, V.E.~Fortov, B.~Militzer, V.B.~Mintsev, and 
  V.Ya.~Ternovoi for stimulating discussions and for providing us with their 
  data. This work was supported by the Deutsche Forschungsgemeinschaft within the 
  SFB~652 and the grant mvp00006 of the supercomputing center HLRN.
%\end{acknowledgements}

%\newpage
%%%%%%%%%%%%%%%%%%%%%%%%%%%%%%% References

\end{document}